\documentclass{article}  
\usepackage{lajolla2006}
\usepackage{graphicx}
\frompage{000} \topage{000}                                              
\def\rightmark{Resonance Production in Heavy Ion Collisions}
\def\leftmark{S. Vogel et al.}
 
\title{Resonance Production in Heavy Ion Collisions - what can we learn from RHIC?} 
\authors{
{S. Vogel$^1$ and M. Bleicher$^{1}$ %
}\\[2.812mm]
{\normalsize
\hspace*{-8pt}$^1$ Institut f\"ur Theoretische Physik, Johann Wolfgang Goethe-Universt\"at\\ 
Max-von-Laue-Str. 1, 60438 Frankfurt am Main, Germany\\[0.2ex] 
}}
 
\abstract{In these proceedings we concentrate on the refeeding and rescattering probability of hadronic resonances. We discuss the probability to form resonances in binary baryon-meson or meson-meson collisions as a function of time for various resonances using a transport model approach (UrQMD). We give an estimate of the re-feeding probability using a simplified thermal approach and discuss the relevance for the resonance/non-resonance ratio measured by STAR.}

 
\begin{document}
 
\maketitle
\setcounter{page}{1}

\section{Introduction}\label{intro}

In order to understand the properties of matter formed in a ultrarelativistic nucleus nucleus collision it is of importance to understand the production and absorption processes of resonant states created in those collisions.
Especially the RHIC program has measured a great amount of data on resonances \cite{Zhang:2004bt,Xu:2001cr,Adler:2002sw,Markert:2004xx,Markert:2003rw,Markert:2002xi,Fachini:2003dx,Fachini:2003mc,Fachini:2004jx,Adler:2004zn,Adams:2004ep}, which is not in line with thermal model estimates \cite{Braun-Munzinger:1995bp,Braun-Munzinger:2001ip,Schweda:2004kd}. Therefore the investigation of the properties and dynamics of resonant structures in a hadronic medium within different approaches is necessary and has been carried out recently \cite{Johnson:1999fv,Soff:2000ae,Torrieri:2001ue,Torrieri:2001tg,Bleicher:2002dm,Torrieri:2002jp,Bleicher:2002rx,Bleicher:2003ij,Torrieri:2004zz,Vogel:2005qr,Vogel:2005pd}.
However, theoretically there are still unsolved problems concerning the regeneration (refeeding) and rescattering processes. It es expected that the inclusion of ``pseudo-elastic'' interactions between chemical and kinetic decoupling might solve some discrepancies of statistical models for resonance yields at high energies.
In the present study we focus on regeneration and rescattering processes in a transport model approach. In Section \ref{rescattering} we concentrate on the rescattering process and discuss the difference of di-leptonic decay channels and di-hadronic decay channels.
In the next section we discuss regeneration effects using a simplified thermal ansatz and also a hadron transport model in order to check for differences in the production channels of those resonances. Finally we conclude with a summary.\\

\begin{figure}[t]
\vspace*{0cm}
\insertplot{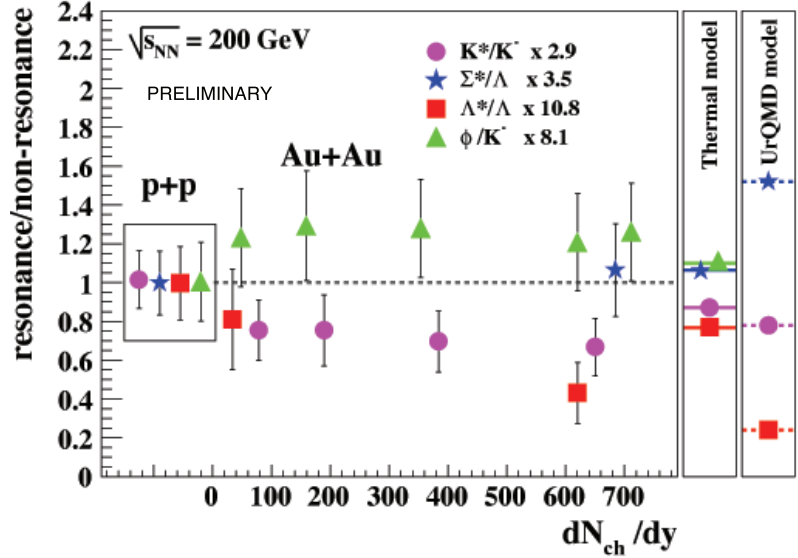}
\vspace*{-.2cm}
\caption[]{(Preliminary) experimental data \cite{christina} for the resonance to non-resonance ratio as a function of centrality ($dN_{ch}/dy$) in Au+Au collison at $\sqrt{s}=200~AGeV$. One observes a decrease of the $K^*/K^-$ and the $\Lambda^*/\Lambda$ ratio, but no change in the $\Phi/K^-$ or the $\Sigma^*/\Lambda$ ratio with respect to the proton proton data.}
\label{expdata}
\end{figure}

Recent experimental data related to rescattering and regeneration processes are shown in Figure \ref{expdata}.
Depicted are resonance/non-resonance ratios for various hadron species as a function of $dN_{ch}/dy$, i.e. charged particles. More central collisions are to the right, proton + proton data points are plotted for comparison to the left. Please note that the curves are scaled such that the proton + proton data normalize to 1.

One observes a decrease of the $K^*/K^-$ and the $\Lambda^*/\Lambda$ ratio, but no change in the $\Phi/K^-$ or the $\Sigma^*/\Lambda$ ratio with respect to the proton proton data. This indicates that not only rescattering effects of resonances \cite{Bleicher:2002dm} have to be taken into account, but refeeding effects as well. 
It has been speculated \cite{christina} that a cross section ordering of the production channels is the reason for this unexpected behaviour. According to the experimental data the production cross sections should obey the relation $\sigma(\Lambda+\pi \rightarrow \Sigma^*) > \sigma(K+\pi \rightarrow K^*) > \sigma(\Sigma+\pi \rightarrow \Lambda^*)$. In fact this feature in observed in the present hadronic transport model with the cross section being $\sigma(\Lambda+X) \rightarrow Y \sim 50mb$, $\sigma(K+X) \rightarrow Y \sim 40mb$ and $\sigma(\Sigma+X) \rightarrow Y \sim 35mb$. Further investigations about the specific production channels are in order.

\section{Rescattering of resonances}\label{rescattering}
In order to adress the topic described above we apply the UrQMD model. It is a non-equilibrium
transport approach based on the covariant propagation of hadrons and strings. All
cross sections are calculated by the
principle of detailed balance or are fitted to available data. 
The model allows to study the full
space time evolution of all hadrons, resonances and their decay products.
This permits to explore the emission patterns of resonances 
in detail and to gain insight of their origins.
For further details about the model the
reader is referred to \cite{Bass:1998ca,Bleicher:1999xi}.

Experimentally, the identification of resonances proceeds via
the reconstruction of the invariant mass distribution (e.g. of charged pions) for
each event. Then, an invariant mass distribution of mixed events is generated (here the
particle pairs are uncorrelated by definition) and subtracted from the mass 
distribution of the correlated events. As a result one obtains the
mass distributions and yields (after all experimental corrections) of
the resonances by fitting the resulting distribution with a suitable 
function (usually a Breit-Wigner distribution peaked around the pole mass). 
For more informations the reader is referred to \cite{Adams:2004ep}.

\begin{figure}[htb]
\vspace*{0cm}
\insertplot{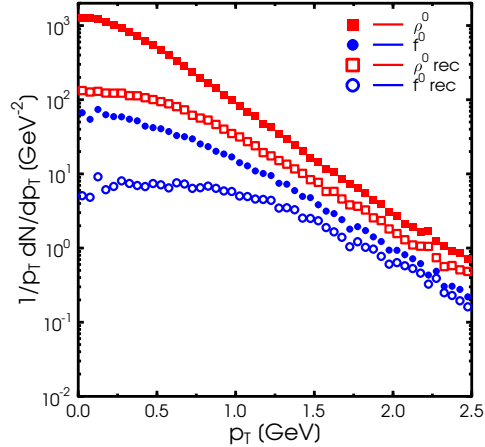}
\vspace*{-.2cm}
\caption[]{Transverse momentum spectrum of $\rho^0$ and $f^0$ mesons for central Au+Au collisions at $\sqrt{s} = 200$AGeV. One observes that the resonances reconstructable via di-hadron correlations (open symbols) differ strongly from the distribution of all decayed resonances (filled symbols)
}
\label{pt_spectra}
\end{figure}

For the model calculation of the resonances, we employ a different 
method to extract the resonances. We follow the decay products of each 
decaying resonance (the daughter particles). If any of the daughter hadrons
rescatters, the signal of this resonance is lost. If the daughter particles
do not rescatter in the further evolution of the system, the resonance is counted
as ``reconstructable''. Note that all decaying resonances are dubbed 
with the term ``all decayed''. These resonances
are reconstructable by an invariant mass analysis of di-leptons (after multiplication with
the respective branching ratio $\Gamma (R\rightarrow e^+e^-)$). 
The advantage of the presently employed method is that it allows 
to trace back the origin of each individual resonance to study their spatial and temporal
emission pattern.
As depicted in Fig. \ref{pt_spectra} one observes a huge difference at low $p_T$ for $\rho^0$ and $f^0$ mesons, which vanishes at higher $p_T$.
This is in line with the rescattering picture, since particles with a large transverse momentum escape the interaction zone more quickly and therefore reduce the chance for their decay prodcuts to rescatter. 
The distribution of all decayed resonances is what one would expect for the decay products of an electromagnetical decay (i.e. leptons). That is because they escape the collision undisturbed and do not undergo rescattering processes.

\section{Regeneration of resonances}

\begin{figure}[htb]
\vspace*{0cm}
\insertplot{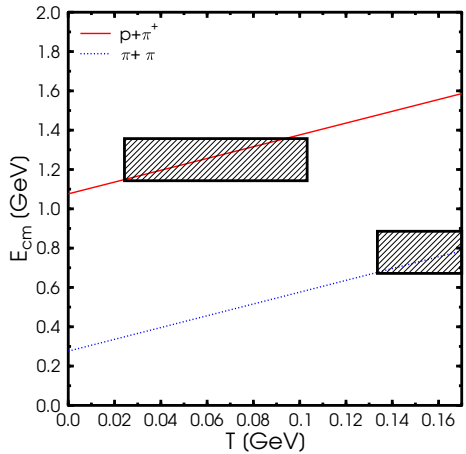}
\vspace*{-.2cm}
\caption[]{$\sqrt{s} \approx m_1 + m_2 +2 \left( \frac{3}{2} T \right)$ as for $\pi\pi \rightarrow \rho$ ($m_1 = m_2 = 138 {\rm MeV}$) and $p\pi \rightarrow \Delta$ ($m_1 = 138 {\rm MeV}, m_2 = 938 {\rm MeV}$) reactions. The striped bars depict the mass region of the $\Delta$ baryons (1232~ {\rm MeV}) and the $\rho$ meson (770 {\rm MeV}), both with a width of roughly 100 {\rm MeV}. One observes, that the $\Delta$ baryon can be produced at temperatures down to 20 {\rm MeV}, whereas the $\rho$ meson can only be produced at temperatures, which are in the order of the kinetic decoupling temperature or above.
}
\label{thermalansatz}
\end{figure}

After having discussed rescattering effects let us now look at refeeding effects.
In order to address the refeeding process let us first get an erstimate using thermodynamics.
Fig. \ref{thermalansatz} depicts the center of mass energy as a function of temperature using the relation 

\begin{eqnarray}
\sqrt{s} \approx m_1 + m_2 +2 \left( \frac{3}{2} T \right). \nonumber
\end{eqnarray}

That equation means that the available energy for particle production (in that case resonances) is the mass of the two colliding particles and two times the thermal energy which should be roughly fulfilled in a thermalized system.

The full line in Fig. \ref{thermalansatz} depicts the above relation with $m_1 = 138~$MeV and $m_2 = 938~$MeV, which is the pion and the proton mass. The dashed line depicts the equation with $m_1 = 138~$MeV and $m_2 = 138~$MeV, which is the pion mass. The striped bars show the mass where the corresponding resonances lie, that is at $E_{cm}$ = 1232~MeV for the $\Delta$ baryon and $E_{cm}$ = 770~MeV for the $\rho$ meson. One observes that the production of $\rho$ mesons is possible only down to temperatures like 130~MeV, whereas the production of the $\Delta$ baryon is principally possible down to temperatures like 20~MeV.

Using a a blastwave model ansatz one obtains a kinetic decoupling temperature of about 90~MeV \cite{Adams:2003xp}, thus $\rho$ mesons cannot be regenerated after kinetic decoupling, whereas $\Delta$ baryons can be regenerated throughout the whole collision.

\begin{figure}[htb]
\vspace*{0cm}
\insertplot{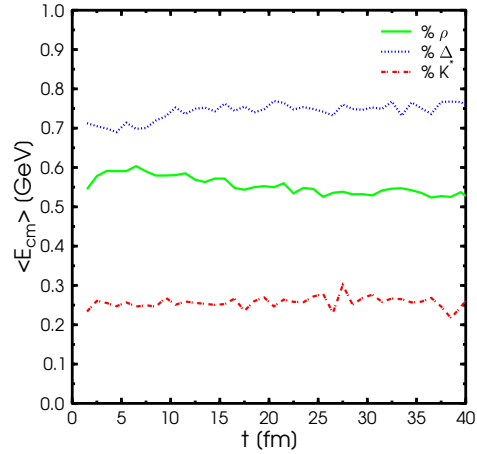}
\vspace*{-.2cm}
\caption[]{Probability to form a resonance in a binary hadron hadron collision as a function of time. Dotted line depicts the probability to produce a $\Delta$ in a binary $p+\pi$ collison.
The full line depicts the probability to form a $\rho$ meson in a binary $\pi+\pi$ collision and the dashed line shows the probability to produce a $K^*$ meson in a $K+\pi$ collision. One observes that the production probability for $\Delta$ baryons is much higher than those for the two other resonances.
}
\label{prob}
\end{figure}

Using a hadronic transport model and taking into account the production and absorption effects dynamically one can trace the binary hadron + hadron collisions and check for resonance production explicitely. Fig. \ref{prob} shows the probability to form a resonance in such a binary hadron + hadron collision ($p+\pi \rightarrow \Delta$ (dotted line), $\pi+\pi \rightarrow \rho$ (full line) and $K+\pi \rightarrow K^*$ (dashed line)) as a function of collision time. 
One observes that in case of a $p+\pi$ collision the chance to actually form a $\Delta$ baryon is higher (70-75\%) than the chance to produce a $\rho$ meson in a $\pi+\pi$ collision (55-60\%) or to produce a $K^*$ meson in a $K+\pi$ collision (25-30\%).

\section{Conclusions}\label{concl}
In summary, the production and rescattering processes of resonances in heavy ion collisions have been studied within a hadronic transport model (UrQMD). It has been shown that rescattering effects are important when comparing di-leptonic and hadronic decay channels. A thermal ansatz has been discussed to estimate the refeeding probability in heavy ion collisions. Probabilities of resonance production in binary hadron + hadron reactions have been studied as a function of collision time. We observed that the probability to create a $\Delta$ baryon in a $p+\pi$ collison is higher than the probability to create a $\rho$ meson in a $\pi+\pi$ collision or a $K^*$ meson in a $K+\pi$ collision. Further studies are ongoing to check the importance of cross section dependences.
 
\section*{Acknowledgments}
This work has been supported by GSI, DFG and BMBF. The authors thank the organizers for a great workshop in La Jolla. Fruitful discussions with Christina Markert, Sevil Salur and Horst St\"ocker are acknowledged.
The computational resources have been provided by the Center for Scientific Computing (CSC) at Frankfurt, Main.
 
\vfill\eject
\end{document}